\documentclass[twocolumn,showpacs,preprintnumbers,amsmath,amssymb]{revtex4}

\usepackage{graphicx}
\usepackage{dcolumn}
\usepackage{bm}

\newcommand{\dhd}{{\textstyle d}
\lower.03ex\hbox{\kern-0.38em$^{\scriptstyle-}$}\kern-0.05em{}} 
\newcommand{\dbar}{{\textstyle \delta}
\lower.03ex\hbox{\kern-0.38em$^{\scriptstyle-}$}\kern-0.05em{}}  

\begin{document}

\preprint{JLAB-THY-02-60}

\title{Scattering of color dipoles: from low to high energies}
\author{Alexander Babansky}
 \email{babansky@jlab.org}
\author{Ian Balitsky}
 \email{balitsky@jlab.org}
\affiliation{
Theory Group, Jefferson Lab, 12000 Jefferson Ave., Newport News, VA 23606\\
and\\
Phys. Dept., Old Dominion Univ., Hampton Blvd,  Norfolk, VA 23529}

\begin{abstract}
A dipole-dipole scattering amplitude is calculated exactly in the first two
orders of perturbation theory.  This amplitude is an analytic function of the 
relative energy and the dipoles' sizes. The cross section of the 
dipole-dipole scattering approaches the high-energy BFKL asymptotics
starting from a relatively large rapidity $\sim 5$.
\end{abstract}

                              
\maketitle

\section{\label{sec:intro}Introduction}
 
It is well known that the effective degrees of freedom at high energies are
so-called Wilson lines - infinite gauge factors ordered along the velocities of
the colliding particles (for a review, see Ref. \onlinecite{mobzor}).
Let us consider the classical example of the scattering of virtual photons in
QCD. The photons decompose into quark-antiquark
pairs which interact by exchanging gluons. At high energies, quarks move
very fast so that their propagation in the background fields of the exchange
gluons is reduced to the gauge factor of $U^n(x_{\perp})$ \cite{nacht}
\begin{equation}
U^n(x_{\perp})=Pe^{ig\int_{-\infty}^{\infty}du n_{\mu}A^{\mu}(un+x_{\perp})}
\label{intro1}
\end{equation}
ordered along the classical trajectory
of the particle, i.e. a straight line collinear to the velocity $n$. 
Here $x_{\perp}$ is the transverse position (an impact parameter) of the fast 
quark, which does not change in the collision.
The propagation of a quark-antiquark pair is described by the 
color dipole \cite{kop} formed from the two Wilson lines
\begin{equation}
W^n(x_{\perp},y_{\perp})={\rm Tr}\{U^n(x_{\perp})U^{n\dagger}(y_{\perp})\}
\label{intro2}
\end{equation} 
where $x_{\perp}$ and 
$y_{\perp}$ are the transverse positions (impact parameters) of quark and 
antiquark.
The high-energy $\gamma^\ast\gamma^\ast$ scattering reduces then to the 
dipole-dipole 
amplitude integrated over dipoles' sizes and separations (see Fig.
\ref{fig:dip0}):  
\begin{figure}
\includegraphics{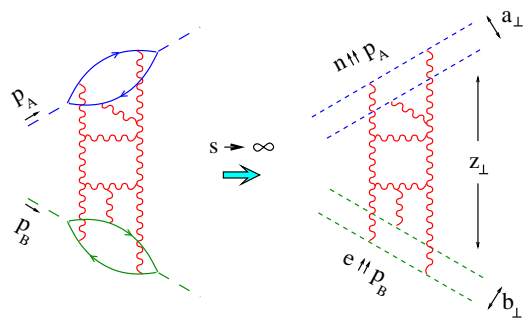}
\caption{\label{fig:dip0} High-energy scattering as a dipole-dipole amplitude.
Wavy lines are gluons and dashed lines are Wilson gauge factors (\ref{intro1})}.
\end{figure}
\begin{equation}
A(s)~=~{s\over 2}\int d^2a_{\perp}d^2b_{\perp}~I^A(a_{\perp})I^B(b_{\perp})
T(a_{\perp},b_{\perp};s)
\label{intro3}
\end{equation}
where $I^A(a_{\perp})$ is the so-called ``impact factor'' (see Appendix A)
and $T(a_{\perp},b_{\perp};s)$ is the dipole-dipole 
scattering amplitude:
\begin{eqnarray}
&&T(a_{\perp},b_{\perp};s)~=\label{intro4}\\
&&-i\!\int\! d^2z_{\perp} 
\langle W^n(a_{\perp}+z_{\perp},z_{\perp})~
W^e(b_{\perp},0_{\perp})\rangle
~.
\nonumber
\end{eqnarray}
Here $e$ is the unit vector in the direction of the motion of the second
virtual photon. The collision energy is related to the relative 
rapidity $\eta$
($\equiv$ angle between the dipoles) 
\begin{equation}
{p_A\cdot p_B\over \sqrt{p_A^2p_B^2}}=n\cdot e=\cosh\eta
~.
\label{rapeta}
\end{equation}
The energy dependence of the $\gamma^\ast\gamma^\ast$ amplitude is 
governed by the evolution of the dipoles with respect to the slope 
of Wilson lines determined by $\eta$ \cite{npb96,prdpl}.

We see that the dipole-dipole amplitude is an ``elementary'' high-energy
scattering process. Moreover, in theories without asymptotic particle states
(like N=4 SYM) the dipole-dipole scattering is the only way to access the
high-energy behavior of amplitudes. Also, there have been many attempts to 
estimate the high-energy behavior of hadron-hadron amplitudes using  
non-perturbative models for the dipole-dipole scattering (see Ref.
\onlinecite{dosch} for a review). For hadrons, the impact factors cannot be 
calculated in pQCD; however, they can be related to the hadron wave
functions. 

The usual phenomenological approach to dipole-dipole scattering in pQCD is to
take the light-like dipoles. However, matrix elements of the light-like 
Wilson operators contain longitudinal divergencies. In the LLA we cut these
divergencies ``by hand''\cite{mu94,nn}. This prescription - take the light-like

 dipoles and cut the divergencies by
hand -  seems to avoid ambiguities in the next-to-leading
order, too\cite{balbel}.  However, to go beyond perturbation theory
one has to consider the scattering of the off-light-cone dipoles.  
The dipole-dipole amplitude is then a function of the dipole separations 
and the angle $\theta$ ($\sim$  relative rapidity $\eta$) between the
dipoles.  If this amplitude is an analytic function of $\theta$, one can 
calculate the dipole-dipole scattering (a correlation function of two infinite
rectangular Wilson loops) in the Euclidean region and continue to the
Minkowski space. 
In the recent series of papers \cite{zahed,pesch}
the dipole-dipole scattering 
was calculated in the Euclidean space using the AdS-CFT correspondence and
continued analytically as a function of the angle $\theta$ to the Minkowski 
space, where $\theta\rightarrow i\eta$.  Another example of such analytic
continuation is the calculation of the instanton-induced dipole-dipole
ampltitude \cite{dodik}.

To be able to obtain the high-energy behavior from the Euclidean amplitudes
it is crucial to verify that the dipole-dipole amplitude is an analytic 
function of $\theta$.
In Ref. \onlinecite{meggio} this statement was proved for the correlation
function of two Wilson rectangles with a finite
longitudinal length  ${\cal T}$ (see also Ref. \onlinecite{pirner}). 
At finite${\cal T}$, the amplitude is an analytic function of the angle 
$\theta$
between the rectangles.  In the limit ${\cal T}\rightarrow\infty$, 
one recovers the dipole-dipole  scattering
amplitude. However, whether the ${\cal T}\rightarrow\infty$ limit 
commutes with the analytic continutation from the Euclidean to Minkowski 
space is an open question. We demonstrate with an explicit
calculation in the first two orders of perturbation theory that 
the dipole-dipole amplitude is, indeed an analytic function 
of the angle between the dipoles (as well as of the dipoles' sizes). 
This is the first important result of our paper.  We see that the analyticity 
survives the  ${\cal T}\rightarrow\infty$ limit in the first two orders 
of perturbation theory, and presumably it happens in all higher orders 
as well so that the dipole-dipole amplitude appears to be an analytic 
function of $\theta$.

The second result of our paper concerns the rate at which the dipole-dipole
amplitude approaches the BFKL asymptotics. At sufficiently high energies,
the cross section for the ``unpolarized'' dipole-dipole scattering
is given by the BFKL formula 
\begin{eqnarray}
\sigma(a,b;\eta)\sim{\alpha_s^2\over ab}\int d\nu \Big({a\over b}\Big)^{2i\nu}
\eta^{{\alpha_s\over\pi}N_c\chi(\nu)}
\label{bfkla}
\end{eqnarray}
where $\chi(\nu)=-\Re\psi(1/2+i\nu)-C$ is the position 
of the ``hard pomeron''\cite{bfkl} 
(for a review, see Ref. \onlinecite{lobzor}). The asymptotics of the 
$\gamma^\ast\gamma^\ast$ scattering is then given by Eq. (\ref{intro3}).
The formula (\ref{bfkla}) is obtained in the so-called leading log approximation 
(LLA), when $\alpha_s\ll 1, \alpha_s\ln s\sim \alpha_s\eta\sim 1$.
The main problem with the LLA result (\ref{bfkla}) is that its power
behavior $s^{4{\alpha_s\over\pi}N_c\ln 2}$  violates
the Froissart bound $\ln^2s$ at asymptotically large $s$. Thus, the BFKL 
pomeron gives only the pre-asymptotic behavior at ``moderately high'' 
energies. It is a common belief that the true
asymptotics at $s\rightarrow \infty$ (where $\alpha\ln s\gg 1$)
comes from the unitarization of the BFKL pomeron. 
Despite numerous efforts, the high-energy 
asymptotics satisfying the Froissart bound is still an unsolved 
problem in QCD (for a recent
discussion see Refs. \onlinecite{mcler02,wkov02,kozlevin}).
In this paper, we address a different problem 
(important for mid-energy accelerators) - when
does the LLA asymptotics start making sense  - at 1, 10 , or 100 GeV? 
To get a complete answer to this question one should
calculate the amplitude 
exactly and compare it to the BFKL asymptotics (\ref{bfkla}). 
Since it seems to be an impossible task at present, 
we calculate the dipole-dipole scattering 
in the first two orders of perturbation theory exactly and compare it 
to the asymptotic form.
Our main result is that the asymptotics starts rather late, at $\eta\sim 5$, 
which translates into $\sqrt{s}\sim$=10 GeV for the scattering of dipoles with
the
 $\rho$-meson size $a\sim 0.3 fm$.

The paper is organized as follows. In Sect. 2 we calculate the dipole-dipole
amplitude and the cross section in the leading order of perturbation theory. 
In the second order, we have calculated independently both the dipole-dipole
amplitude and its imaginary part proportional to the cross section. Since the 
structure of the diagrams and the result is much more transparent for the
cross section, in Sect. 3 we present only the results for the 
amplitude in the second order and leave the details until Sect. 5, where
we give the detailed calculation of the dipole-dipole cross section.   
We discuss the effects due to the running coupling constant in Sect 4. 
 Sect 6 contains the numerical estimates of 
the dipole-dipole cross section and Sect. 7 is devoted to the discussion of 
our results and an outlook. The explicit form of the second-order amplitudes 
is given in the Appendix. 

\section{\label{lowest} Dipole-dipole scattering in the Born approximation}

We consider the dipole-dipole scattering amplitude defined as follows
\begin{eqnarray} 
&&T(x_{1\perp},x_{2\perp};x_{3\perp},x_{4\perp};\eta)                          
 \label{triv1}\\
&&=~-i\langle W^n(x_{1\perp},x_{2\perp})
W^e(x_{3\perp},_{4\perp})\rangle
~.
\nonumber
\end{eqnarray}
Strictly speaking, the Wilson lines forming the dipole 
are connected with gauge links at
infinity, so 
\begin{eqnarray} 
&&T(x_{1\perp},x_{2\perp};x_{3\perp},x_{4\perp};\eta)\nonumber\\
&&=~-i\lim_{{\cal T}\rightarrow\infty}
\langle W^n_{\cal T}(x_{1\perp},x_{2\perp})
W^e_{\cal T}(x_{3\perp},x_{4\perp})\rangle
\label{triv2}
\end{eqnarray}
where $W^n_{\cal T}$ is a rectangular Wilson loop 
of longitudinal length ${\cal T}$
\begin{eqnarray} 
&& 
W^n_{\cal T}(x_{1\perp},x_{2\perp})={\rm Tr} \{U_{\cal T}^n(x_{1\perp})
[x_{1\perp}-{{\cal T}\over 2}n,x_{2\perp}-{{\cal T}\over 2}n]
\nonumber\\
&&\times~
U_{\cal T}^{n\dagger}(x_{2\perp})[x_{1\perp}+
{{\cal T}\over 2}n,x_{2\perp}+{{\cal T}\over 2}n]\} 
\label{triv3}
\end{eqnarray}
and $U_{\cal T}^n(x_{1\perp})\equiv 
[{{\cal T}\over 2} n+x_{1\perp},-{{\cal T}\over 2} n+x_{1\perp}]$ 
etc.
Hereafter, we use the notation
\begin{equation}
[x,y]\equiv Pe^{ig\int_0^1 du(x-y)^{\mu}A_{\mu}(ux+(1-u)y)}
\label{pexp}
sec:
\end{equation}
for the straight-line ordered gauge link suspended
between the points $x$ and $y$.

Since the gluons reduce to the pure gauge fields at infinity, 
the precise form of a contour
connecting the points $x_{1\perp}\pm {{\cal T}\over 2}n$ and 
$x_{2\perp}\pm {{\cal T}\over 2}n$ 
does not matter. Moreover, we use Feynman gauge for the gluon propagator,
and in this gauge the links at infinity do not contribute to the amplitude.
Thus, we omit them in what follows to simplify the notations.
  
\subsection{Cross section of the dipole-dipole scattering
and optical theorem for dipoles}

The total cross section of the scattering of the dipole of size $a$
on the dipole of size $b$  may be defined as
\begin{equation}
\sigma(a_{\perp},b_{\perp};\eta)~=~\int\!dz_{\perp}
\varsigma(a_{\perp}+z_{\perp};z_{\perp},b_{\perp};\eta )
\label{opt0}
\end{equation}
where 
\begin{eqnarray}
\varsigma(x_{i\perp}, \eta)&\equiv&
\sum_{X\neq 0}\langle 0|W_{ij}^{n\dagger}(x_{1\perp},x_{2\perp})
W_{kl}^{e\dagger}(x_{3\perp},x_{4\perp})|X\rangle\nonumber\\
&\times&\langle X|W_{ji}^n(x_{1\perp},x_{2\perp})
W_{lk}^e(x_{3\perp},x_{4\perp})|0\rangle
\label{opt1}
\end{eqnarray}
is the total amplitude of the dipole-dipole transition into hadrons. 
Here the summation goes over all the intermediate states except for 
the vacuum and
\begin{eqnarray} 
&& \hspace{-4mm}
W^n_{ij}(x_{1\perp},x_{2\perp})\label{opt2}\\
&&\hspace{-4mm} =~\lim_{{\cal T}\rightarrow \infty}
\Big(U_{\cal T}^n(x_{1\perp})
[x_{1\perp}-{{\cal T}\over 2}n,x_{2\perp}-{{\cal T}\over 2}n]
U_{\cal T}^{n\dagger}(x_{2\perp}) \Big)_{ij}
~.
\nonumber
\end{eqnarray}
If we include the intermediate vacuum state in the r.h.s. of Eq. (\ref{opt1}),
we get $N_c^2$ due to the completeness relation. 
Separating the non-interacting contribution in a usual way 
\begin{equation}
\langle 0|W^n(x_{1\perp},x_{2\perp})
W^e(x_{3\perp},x_{4\perp})|0\rangle=N_c^2+iT(x_i,\eta) \,
\label{opt3}
\end{equation}
we get the ``optical theorem'' for the dipole-dipole scattering:
\begin{equation}
\Im T(x_{1\perp},x_{2\perp};x_{3\perp},x_{4\perp};\eta)
=~{1\over 2}\varsigma(x_{1\perp},x_{2\perp};x_{3\perp},x_{4\perp};\eta)
~.
\label{opt4}
\end{equation}

 It is worth noting that the high-energy $\gamma^\ast\gamma^\ast$ cross section
reduces to the dipole cross section (\ref{opt1}) integrated with impact factors,
similar to the Eq. (\ref{intro3}) for the amplitude:
\begin{equation}
A(s)~=~{1\over 4}\int d^2a_{\perp}d^2b_{\perp}~I^A(a_{\perp})I^B(b_{\perp})
\sigma(a_{\perp},b_{\perp};\eta)
\label{opt5}
\end{equation}

We will calculate the dipole-dipole cross section in two ways: directly as the
r.h.s. of Eq. (\ref{opt1}) or via the optical theorem as the imaginary part 
of eq. (\ref{triv1}). 

\subsection{Lowest order dipole-dipole amplitude} 

In the leading order of perturbation theory, the correlation function of two
Wilson-line operators is proportional to the (massless) two-dimensional 
propagator:
\begin{eqnarray} 
&&\langle U^n(x_{\perp})U^e(y_{\perp})~=~
-\langle U^n(x_{\perp})U^{e\dagger}(y_{\perp})\rangle\nonumber\\
&&=~ig^2\coth\eta
\int\!{\dhd^2 k_{\perp}\over k_{\perp}^2}
~e^{i(k,x-y)_{\perp}}
\label{lodipa1}
\end{eqnarray}
where $(a,b)_{\perp}$ denotes the (positive) scalar product of two-dimensional 
vectors $\vec{a}_{\perp}$ and $\vec{b}_{\perp}$. (For brevity, we use the 
$\hbar$-inspired notations $\dhd^n k\equiv{d^n k\over(2\pi)^n}$ and
$\dbar^{(n)}(k)\equiv(2\pi)^n\delta^{(n)}(k)$). 
 
\begin{figure}
\includegraphics{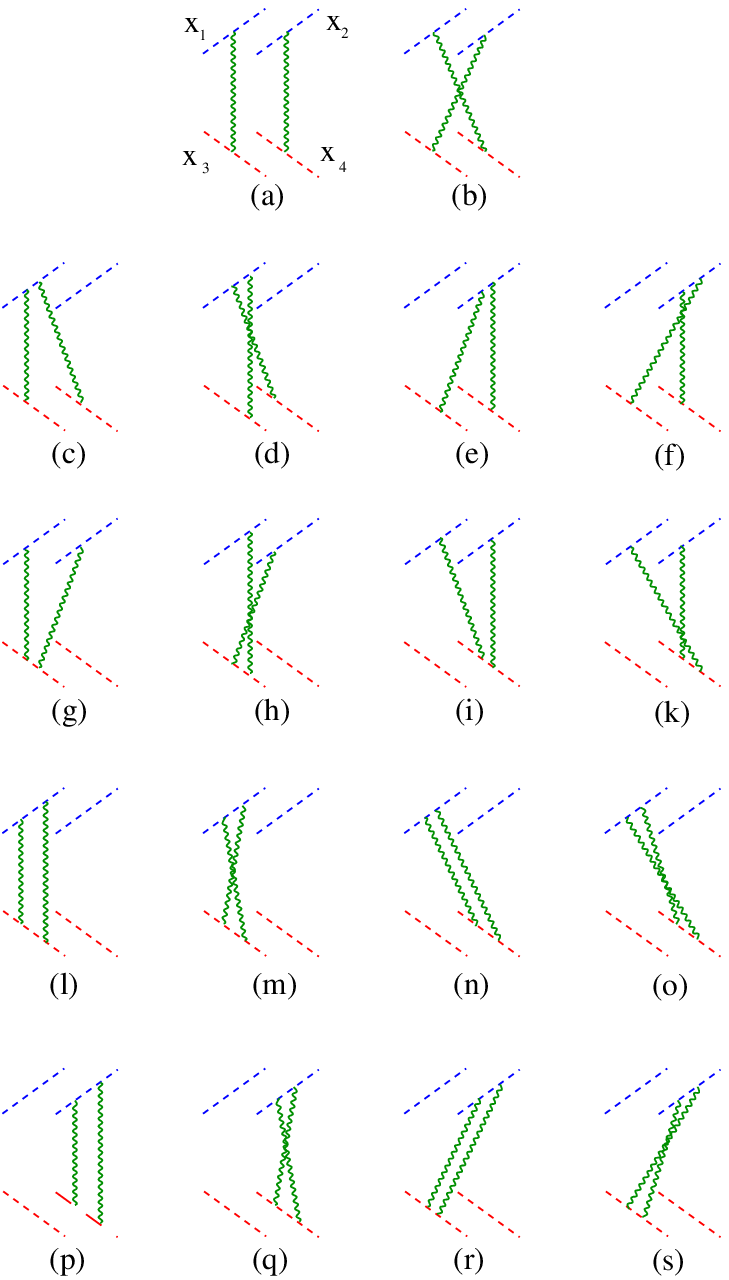}
\caption{\label{fig:dip1} Dipole-dipole scattering in the lowest order
of perturbation theory}
\end{figure}

The relevant Feynman diagrams are shown in Fig. \ref{fig:dip1}.
The contribution of these diargams has the form:

\begin{eqnarray} 
&&\hspace{0mm} 
T(x_{1\perp},x_{2\perp};x_{3\perp},x_{4\perp};\eta) 
\label{formula0} \\ 
&&\hspace{0mm} 
=~ig^4\coth^2\eta ~{N_c^2-1\over 8}
\int\!{d^2 k_{1\perp}d^2 k_{2\perp}\over 16\pi^4k_{1\perp}^2k_{2\perp}^2} 
\nonumber\\ 
&&\hspace{0mm}\times~
(e^{-i(k_1,x_1)_{\perp}}-e^{-i(k_1,x_2)_{\perp}}) 
(e^{i(k_1,x_3)_{\perp}}-e^{i(k_1,x_4)_{\perp}})
 \nonumber\\ 
&&\hspace{0mm}\times
~(e^{i(k_2,x_1)_{\perp}}-e^{i(k_2,x_2)_{\perp}})  
(e^{-i(k_2,x_3)_{\perp}}-e^{-i(k_2,x_4)_{\perp}}) 
\nonumber
\end{eqnarray} 
The integral over the transverse momenta can be performed explicitly, 
resulting in 
\begin{eqnarray}
&&
T(x_{1\perp},x_{2\perp};x_{3\perp},x_{4\perp};\eta)\nonumber\\
&&=~i{N_c^2-1\over 8}\alpha_s^2\left(\ln{x^2_{13}x^2_{24}\over
x^2_{23}x^2_{14}}\right)^2 \coth^2\eta
\label{trivorder}
\end{eqnarray}
where $x_{12}\equiv x_{1\perp}-x_{2\perp}$ etc.
 (cf. Ref.
\onlinecite{kozlevin})
 
\subsection{Dipole-dipole cross section in the Born approximation}

As we mentioned above, the dipole-dipole cross section can be obtained in two
ways: via the optical theorem as the imaginary part of Eq. (\ref{trivorder})
or, directly, as the r.h.s. of Eq. (\ref{opt3}). 
Since the amplitude in the lowest order in $\alpha_s$ is purely imaginary, 
the ``unintegrated cross section'' (\ref{opt1}) is 
\begin{eqnarray}
&&\hspace{0mm}
\varsigma(x_{i\perp};\eta)~=~
g^4{N_c^2-1\over 4}\coth^2\eta~\left(\ln{x^2_{13}x^2_{24}\over
x^2_{23}x^2_{14}}\right)^2~.
\label{trivordersech}
\end{eqnarray}
For future use, it is instructive to calculate $\varsigma(x_{i\perp},\eta)$
directly as the r.h.s. of Eq. (\ref{opt1}). In this case, Feynman
rules for the calculation of the cross sections can be reproduced by a
functional integral over the double set of 
fields (see e.g. \cite{keld}): (+) to the right of
the cut and (-) to the left with the propagators
\begin{eqnarray}
A^{a\mu}_{\pm}(x)A^{b\nu}_{\pm}(y)&=&\!\int\! {dk\over 16\pi^4} e^{-ik(x-y)}
{ig^{\mu\nu}\delta^{ab}\over \mp k^2-i\epsilon}\label{props}\\
 A^{a\mu}_{-}(x)A^{b\nu}_{+}(y)&=&-\!\int\!{dk\over 16\pi^4} e^{-ik(x-y)}
g^{\mu\nu}\delta^{ab}2\pi\delta(k^2)\theta(k_0)
\nonumber
\end{eqnarray}
These correspond to the Cutkovsky rules
for the cross sections. 

\begin{figure}
\includegraphics{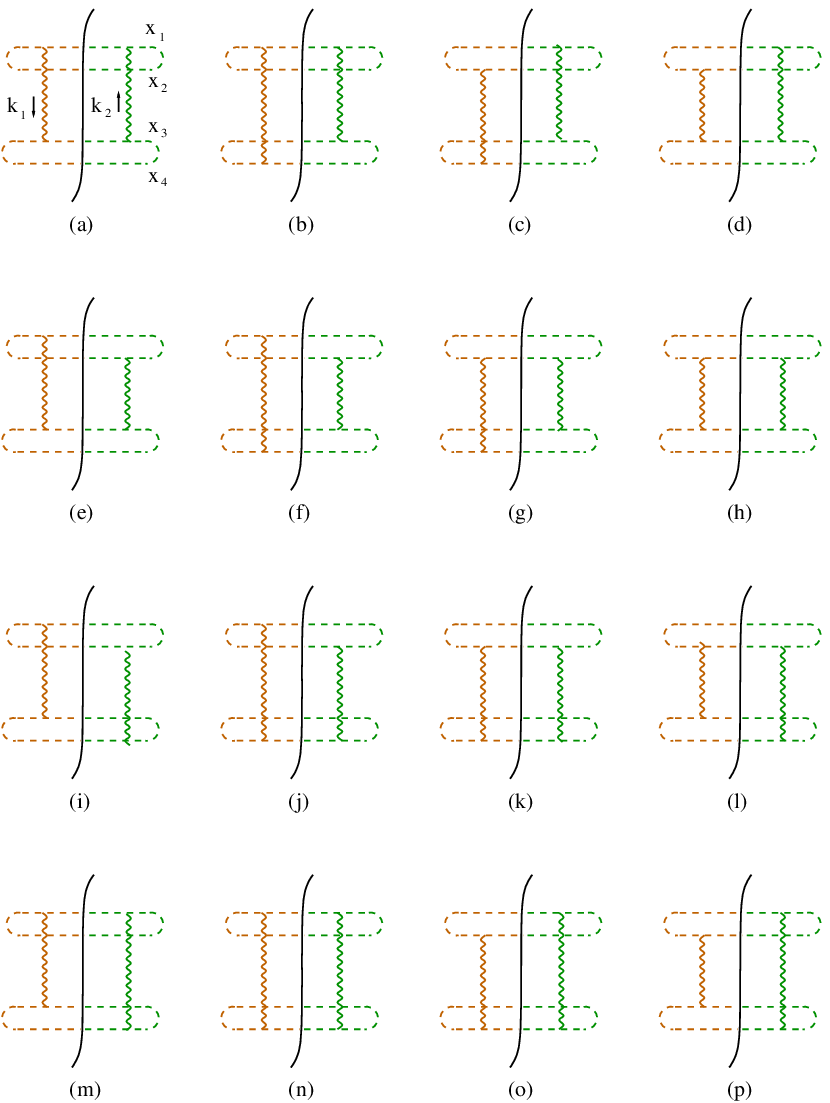}
\caption{\label{fig:dip2} Cross section of the dipole-dipole scattering in the
lowest order of perturbation theory}
\end{figure}

In these notations, the dipole-dipole cross section reads (cf.
\cite{difope}) 
\begin{eqnarray}
\hspace{-5mm}
\varsigma(x_{i\perp};\eta)&=&
\langle {\rm Tr}\{ W^n_{(-)}(x_{1\perp},x_{2\perp})
W^{n}_{(+)}(x_{1\perp},x_{2\perp})\}
\nonumber\\
&\times&~{\rm Tr}\{
W^e_{(-)}(x_{3\perp},x_{4\perp})W^e_{(+)}(x_{3\perp},x_{4\perp})\}
\rangle
\label{flaxz7} 
\end{eqnarray}
The relevant diagram is shown in Fig. (\ref{fig:dip2}).
To emphasize that here we have only the gauge links at ${\cal T}=-\infty$ we
draw them  explicitly.

The leading-order correlation function of Wilson lines in $(+)$ sector 
is given by Eq. (\ref{lodipa1}) and in $(-)$ sector by Eq. (\ref{lodipa1})
with a different sign
\begin{eqnarray} 
&&\langle U_{(+)}^n(x_{\perp})U_{(+)}^e(y_{\perp})~=~
-\langle U_{(+)}^n(x_{\perp})U_{(+)}^{e\dagger}(y_{\perp})\rangle\nonumber\\
&&=~-\langle U_{(-)}^n(x_{\perp})U_{(-)}^e(y_{\perp})~=~
\langle U_{(-)}^n(x_{\perp})U_{(-)}^{e\dagger}(y_{\perp})\rangle\nonumber\\
&&=~ig^2\coth\eta
\int\!{d^2 k_{\perp}\over 4\pi^2k_{\perp}^2}
~e^{i(k,x-y)_{\perp}}
\label{lodips4}
\end{eqnarray}
Using Eq. (\ref{lodips4}) it is easy to see that 
\begin{eqnarray}
&&\hspace{0mm}
{4\over N_c^2-1}\varsigma(x_{1\perp},x_{2\perp};x_{3\perp},x_{4\perp};\eta)
\nonumber\\
&&\hspace{0mm}
=~g^4\coth^2\eta
\int\!{d^2 k_{1\perp}d^2 k_{2\perp}\over 16\pi^4k_{1\perp}^2k_{2\perp}^2}
(e^{-i(k_1,x_1)_{\perp}}-e^{-i(k_1,x_2)_{\perp}})
\nonumber\\
&&\hspace{0mm}\times~
(e^{i(k_1,x_3)_{\perp}}-e^{i(k_1,x_4)_{\perp}})
~(e^{i(k_2,x_1)_{\perp}}-e^{i(k_2,x_2)_{\perp}})
\nonumber\\
&&\hspace{0mm}\times~
(e^{-i(k_2,x_3)_{\perp}}-e^{-i(k_2,x_4)_{\perp}})
\label{lodips5}
\end{eqnarray}
where the $2^4=16$  terms correspond to 16 diagrams in Fig. \ref{fig:dip2}).
Integrating over $k_{\perp}$, we reproduce the optical-theorem result 
(\ref{trivordersech}).

\section{Second-order amplitude}

\begin{figure}
\includegraphics{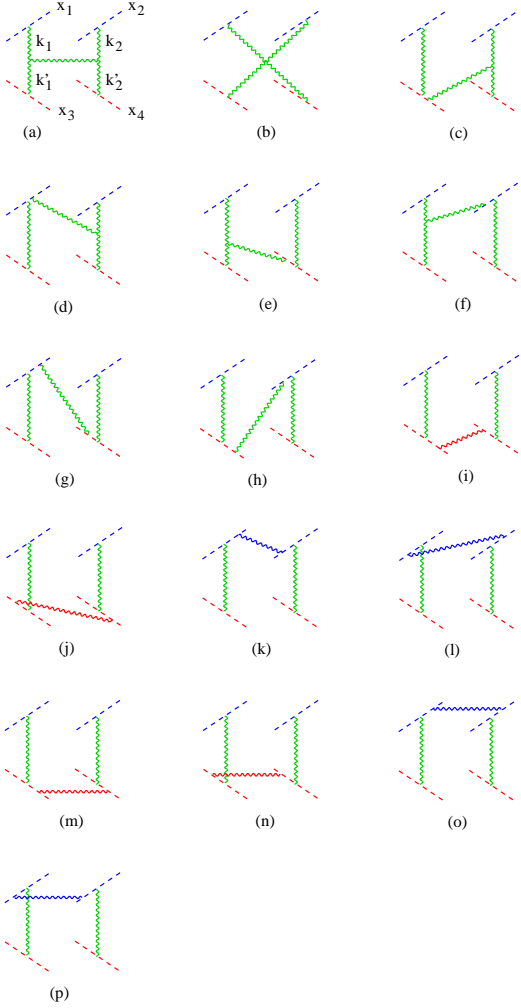}
\caption{\label{fig:dip3} Typical ``connected '' diagrams for the 
dipole-dipole amplitude.}
\end{figure}

In the next order of perturbation theory there are too many diagrams to 
present them all - rougly speaking, we must take the diagrams in
Fig. \ref{fig:dip1} and add an extra gluon line in all possible ways. 
Let us consider the diagram in Fig.  \ref{fig:dip1}a. If we insert the gluon 
line in all possible ways to connect the left and the right parts of the
diagram  in Fig. \ref{fig:dip1}a, we get the diagrams shown 
in Fig. \ref{fig:dip3}.

\begin{figure}
\includegraphics{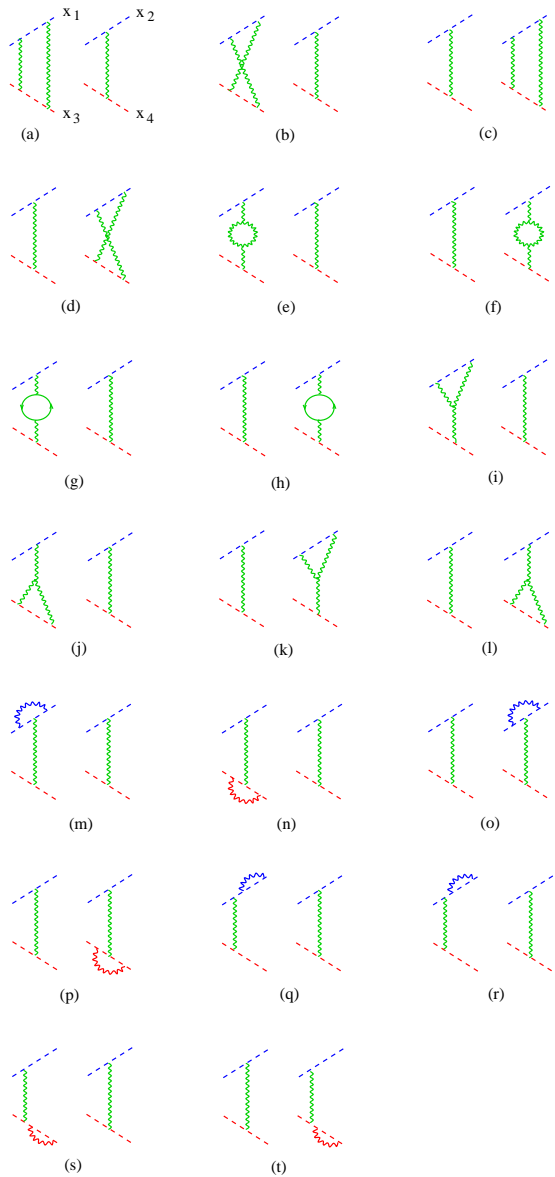}
\caption{\label{fig:dip4}`'Disconnected'' diagrams
for the dipole-dipole amplitude.} \end{figure}

Also, we can insert a gluon line in the left or right part only and get
 the
``disconnected'' diagrams  shown in Fig. \ref{fig:dip4}.

Not all the diagrams for the correlator of two Wilson loops (\ref{triv1}) 
contribute to the dipole-dipole scattering amplitude.  There are two
exceptional classes of diagrams proportional to the total time of the
evolution ${\cal T}$. First, there are those shown in Fig. \ref{fig:dip4}
q-t,  which describe ``mass renormalization'' of Wilson lines. 
Their contribution has the usual form of
$\delta m L$, where $\delta m$ is the self-energy correction and 
$L\simeq 2{\cal T}$ is the perimeter of 
the Wilson loop corresponding to the dipole.
Similarly, the diagrams in Fig. \ref{fig:dip3} m-p contribute to the ``binding
energy'' $\epsilon$ of the dipole and lead to $\epsilon {\cal T}$ terms. 
These two contributions exponentiate so that one obtains:
\begin{eqnarray}
&&\hspace{0mm}
-i\langle 
{\rm Tr}\{U_{\cal T}^n(x_1)U_{\cal T}^{n\dagger}(x_2)\}~
{\rm Tr}\{U_{\cal T}^n(x_3)U_{\cal T}^{n\dagger}(x_4)\}\rangle
\label{fo1}\\
&&\hspace{0mm}=~
e^{-4i\delta m {\cal T}-i(\epsilon_{12}+\epsilon_{34}){\cal T}}
~T(x_{1\perp},x_{2\perp},x_{3\perp},x_{4\perp},\eta;{\cal T})
\nonumber
\end{eqnarray}
where 
$T(x_{i\perp},\eta;{\cal T})$ describes the dipole-dipole scattering 
at the time ${\cal T}$. 
The resulting 
amplitude 
\begin{eqnarray}
&&\hspace{0mm}
T(x_{i\perp},\eta)=
\lim_{{\cal T}\rightarrow\infty}T(x_{i\perp},\eta;{\cal T})
\label{fo2}\\
&&\hspace{0mm}=~\lim_{{\cal T}\rightarrow\infty}
-i\langle 
{\rm Tr}\{U_{\cal T}^n(x_1)U_{\cal T}^{n\dagger}(x_2)\}~
{\rm Tr}\{U_{\cal T}^n(x_3)U_{\cal T}^{n\dagger}(x_4)\}\nonumber\\
&&\hspace{50mm}
\times~e^{4i\delta m {\cal T}+i(\epsilon_{12}+\epsilon_{34}){\cal T}}
\nonumber
\end{eqnarray}
is finite as ${\cal T}\rightarrow\infty$. 
In the second order of perturbation theory, the multiplication in r.h.s. of 
eq. (\ref{fo2}) reduces to the subtraction of the corresponding terms 
($4\delta m {\cal T}$ and $2\epsilon {\cal T}$).
The result is that the diagrams in Fig. \ref{fig:dip3} m-p and
Fig. \ref{fig:dip4} q-t are left out and
in the diagrams in Fig. \ref{fig:dip3} i-l and Fig. \ref{fig:dip4} m-p the 
color factor $-{1\over 2N_c}$ 
is replaced by $-{1\over 2N_c}-c_F=-{N_c\over 2}$.        

Thus, we are left with the diagrams shown in Fig. \ref{fig:dip3}a-j and
\ref{fig:dip4}a-r. Also, there are $\sim 20$ times more diagrams which 
are obtained from the 
addition of an extra gluon to the graphs in Fig. 2b,c,...s. 
The calculation of these diagrams is standard but lengthy.  In Sect.
\ref{crsc} we will present some details of the (independent) 
calculation of the imaginary
part of this amplitude where the structure of diagrams is much more
transparent. Here we will only give the final result for the amplitude and
discuss which classes of diagrams contribute to it.

We have performed calculation of the dipole-dipole amplitude in both 
the Euclidean and Minkowski spaces. In the Euclidean space, this amplitude is
a correlation function of two Wilson rectangles at
angle $\theta$ such as $\cos\theta=n\cdot e$.
In both cases, the $g^6$ amplitude has the form (in 
Euclidean space $\eta=i\theta$):
\begin{eqnarray}
&&\hspace{-3mm}T(x_{i\perp},\eta)
~ =~{ig^6\over 4\pi}{N_c(N_c^2-1)\over 8} \coth^2\eta
\label{mastereqa}\\
&&\hspace{-3mm}\times
\int\!\dhd^2 k_1\dhd^2 k_2
\dhd^2 k'_1 \dhd^2 k'_2~
(e^{-i(k_1,x_1)}-e^{-i(k_1,x_2)})
\nonumber\\
&&\hspace{-3mm}\times~
(e^{-i(k'_1,x_3)}-e^{-i(k'_1,x_4)})
(e^{i(k_2,x_1)}-e^{i(k_2,x_2)})
\nonumber\\
&&\hspace{-3mm}\times~
(e^{i(k'_2,x_3)_{\perp}}-e^{i(k'_2,x_4)_{\perp}})
~\Bigg\{\dbar^2(k_1+k'_1-k_2-k'_2)
\nonumber\\
&&\hspace{-3mm}\times~
\Bigg[{A(k_i;\eta)\over(k_1+k'_1)^6}+
i\pi{(N_c^2-4)\coth\eta\over N^2_c(k_1+k'_1)_{\perp}^2}
\Big({1\over k_1^2{k'}^2_2}+{1\over 
k_2^2{k'}^2_1}
\Big)\Bigg]
\nonumber\\ 
&&\hspace{-3mm}+~\dbar^2(k_1+k'_1)\dbar^2(k_2+k'_2)~
\Bigg[
-\Big(\eta -{i\pi\over 2}+{i\pi\over 2}{N_c^2-4\over N_c^2}\Big)
\nonumber\\
&&\hspace{-3mm}\times~\coth\eta
\int\!\dhd^2 p~\Big\{{k_1^2\over p^2(k_1-p)^2}+
{k_2^2\over p^2(k_2-p)^2}\Big\}\nonumber\\
&&\hspace{-3mm}+~
{1\over 2\pi}\ln {\mu^2\over |k_1||k_2|}
\Big({5\over 3}-{n_f\over 2N_c}\Big)
+{1\over \pi}\ln \mu^2|x_{12}||x_{34}|\Bigg]\Bigg\}
~.
\nonumber
\end{eqnarray}
Here $\mu$ is the normalization point of the $\overline{MS}$ scheme. 
The explicit form of the function $A(k_1,k'_1,k_2,k'_2;\eta)$ 
is given in the Appendix B.
Note that they are analytical functions of $\eta$ and $k_{i\perp}^2$ with
singularities at $\eta\rightarrow 0$ corresponding to the bound state of
two parallel dipoles.

The r.h.s. of Eq. (\ref{mastereqa}) consists of two terms.
The first term comes from
the diagrams of the type shown in Fig. \ref{fig:dip3}. 
(The $N_c^2-4$ term is the odderon
contribution coming from the structure $d^{abc}d^{abc}$).
The first expression of the second term comes from the 
``gluon reggeization'' diagrams in  
Fig. \ref{fig:dip4} a-d, while the second one 
($\sim {5\over 3}-{2n_f\over 3N_c}$) comes from
the gluon and quark self-energy shown in Fig. \ref{fig:dip4} e-h.  
The diagrams in  Fig.\ref{fig:dip4} i-l vanish (cf. Ref. \onlinecite{hquarks}),
while the terms coming from Fig.\ref{fig:dip4} m-p are the pure divergency 
$\sim \int {d^2 k_{\perp}\over k_{\perp}^2}$ which also vanishes in the
framework of dimensional regularization(see the discussion in next
Section). It is easy to see that the apparent divergency at $k+k'\rightarrow 0$
in the gluon reggeizatoon terms cancels with the corresponding contribution to
$A(k_i;\eta)$ (see Eq. \ref{B2}). 

The main conclusion from the Eq. (\ref{mastereqa}) is that the
dipole-dipole amplitude is an analytic function of the angle $\eta$.  
It is easy to check
that the functions $A$ become real as $\eta\rightarrow i\theta$ 
(see the Appendix B) and so does the amplitude $iT(x_{i\perp},i\theta)$.

It is worth noting that in the LLA (as $\eta\rightarrow\infty$) we reproduce
the first iteration of the BFKL kernel
\begin{eqnarray}
&&\hspace{-3mm}T(x_{i\perp},\eta)
~\stackrel{\eta\rightarrow \infty}{\rightarrow}~{ig^6\eta\over
4\pi}{N_c(N_c^2-1)\over 4}  \label{bfklas}\\
&&\hspace{-3mm}\times
\int\!\dhd^2 k_1\dhd^2 k_2
\dhd^2 k'_1 \dhd^2 k'_2~
(e^{-i(k_1,x_1)}-e^{-i(k_1,x_2)})
\nonumber\\
&&\hspace{-3mm}\times~
(e^{-i(k'_1,x_3)}-e^{-i(k'_1,x_4)})
(e^{i(k_2,x_1)}-e^{i(k_2,x_2)})
\nonumber\\
&&\hspace{-3mm}\times~
(e^{i(k'_2,x_3)_{\perp}}-e^{i(k'_2,x_4)_{\perp}})
~\dbar^2(k_1+k'_1-k_2-k'_2)
\nonumber\\
&&\hspace{-3mm}\times~
\Bigg[ -{(k_1-k_2)^2\over k_1^2k_2^2{k'_1}^2{k'_2}^2}
+{k_1^{-2}{k'_2}^{-2}+k_2^{-2}{k'_1}^{-2}
\over {k'_2}^2(k_1+k'_1)^2}
\nonumber\\ 
&&\hspace{-3mm}-~\dbar^2(k_1+k'_1)
\int\!\dhd^2 p~\Big\{{k_1^2\over p^2(k_1-p)^2}+
{k_2^2\over p^2(k_2-p)^2}\Big\}\Bigg]
~.
\nonumber
\end{eqnarray}

\section{\label{sec:ricici}Running coupling constant}

\begin{figure}
\includegraphics{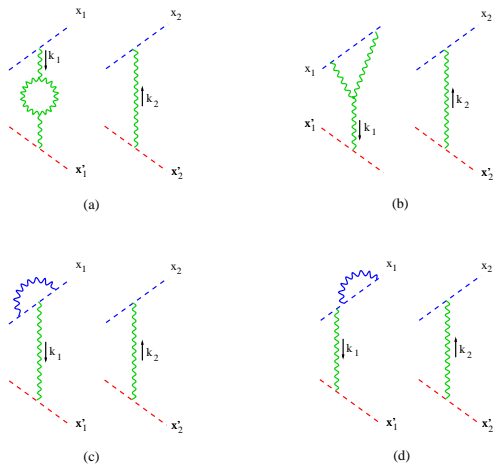}
\caption{\label{fig:dip5} Running coupling constant for the light-like
dipoles}
 \end{figure}

As we shall see below, the diagrams contributing to the renormalization of
coupling constant are somewhat unusual: as in the case of a heavy-quark
potential ${\alpha_s(r)\over r}$, the coefficient 
${11\over 3}N_c-{2n_f\over 3}$ in front of $\ln\mu$ comes from both the 
UV- and the IR-divergent diagrams (cf. \cite{hquarks}).To see how it 
happens, it is instructive to start with the scattering of the light-like
dipoles where the diagrams for the renormalization of coupling constant are
the usual UV-divergent vertices and self-energies shown in Fig. \ref{fig:dip5}.

As we mentioned above, the scattering amplitude for the light-like dipoles is
not a well-defined quantity because of the longitudinal divergencies
corresponding to $\eta=\infty$ coming from the diagrams in Fig. \ref{fig:dip3}
a-h and Fig. \ref{fig:dip4} a-d. If we cut this divergency ``by hand'', we get 
the asymptotical contribution, i.e. Eq. (\ref{bfklas}). Apart from this 
contribution, the only non-zero diagrams are those shown in 
Fig. \ref{fig:dip4} e-h. They lead to the renormalization of coupling 
constant in Eq. (\ref{trivorder}).

Standard computation of these diagrams yields
\begin{equation}
\Big[\Big({5\over 3}N_c-{2n_f\over 3}\Big)+2N_c\Big]~
{g^2\over 16\pi^2}\ln{\mu^2\over k_{1\perp}^2}
\end{equation}
where $\Big({5\over 3}N_c-{2n_f\over 3}\Big)$ comes from the diagram 
in Fig. \ref{fig:dip5}a and the similar diagram with the quark loop
(see Fig. \ref{fig:dip4}o),
while $2N_c$ comes from the diagram in Fig. \ref{fig:dip5}b. 
It turns out that the diagram in Fig.  \ref{fig:dip5}c
vanishes, while the diagram in Fig.  \ref{fig:dip5}d is irrelevant since it
contributes only to the ``mass renormalization'' $\delta m$ of the Wilson 
line (see Eq. (\ref{fo1})).

\begin{figure}
\includegraphics{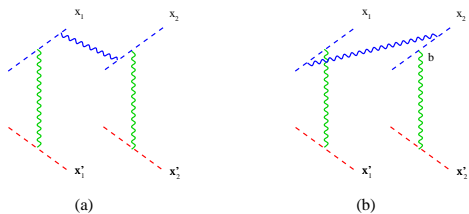}
\caption{\label{fig:dip6} IR divergent diagrams contributing to the 
renormalization of the coupling constant} 
\end{figure}

Surprisingly, the structure of the diagrams contributing to the running coupling
constant is quite different for even slightly off-the-light-cone dipoles.  
In this case the diagram in Fig. \ref{fig:dip5}a is
the same as for the light-like dipoles, the
diagram in Fig. \ref{fig:dip5}b vanishes, and the diagram in Fig.
\ref{fig:dip5}c is a pure divergency of the type $\int {d p_{\perp}\over
p_{\perp}^2}$, which must be set to zero in the dimensional
regularization approach. 
Following Ref. \cite{hquarks} it is convenient to
write down this term as 
\begin{equation}
{g^2N_c\over 8\pi^2}\Big({2\over 4-d}-{2\over d-4}\Big)
\end{equation}
The UV pole ${2\over 4-d}$ together with the pole coming from the diagram 
in Fig. \ref{fig:dip5}a forms 
${2\over 4-d}{g^2\over 16\pi^2}\Big({11\over 3}N_c-{2\over 3}n_f\Big)$,
while the IR pole ${2\over d-4}$ cancels with similar poles in the diagrams in
Fig. \ref{fig:dip6}.

Due to ``cancellation'' between the UV and IR contributions to Fig.
\ref{fig:dip5}c, in order to reproduce the running-coupling
coefficient  $\Big({11\over 3}N_c-{2\over 3}n_f\Big)$
in front of $\ln\mu^2$ in the dimensional regularization, one should take into
account not only the ``usual'' UV divergent diagrams in Fig. \ref{fig:dip5} 
but also the IR-divergent diagrams in Fig. \ref{fig:dip6} (cf.
\cite{diplom}).  As a result the coefficient $\Big({5\over 3}N_c-{2n_f\over
3}\Big)$ in front of ${g^2\over 16\pi^2}\ln\mu^2$ comes from the diagram 
in Fig. \ref{fig:dip5}a, while  ${g^2N_c\over 8\pi^2}\ln\mu^2$ comes from
the diagrams in Fig. \ref{fig:dip6} 
(and all other diagrams in Fig. \ref{fig:dip5} vanish).

This mixture of the IR and UV divergencies could have been a source of a
potential problem.  The cornerstone of the BFKL approach is the assumption
that all the $\ln s$ come from the longitudinal integrations, while the
integrations over the transverse momenta are convergent at $k_{\perp}^2\sim
m^2$, where $m$ is of order of masses (or virtualities) of the colliding
particles. (After the summation of the BFKL ladder we have ``diffusion''
in the transverse momenta leading to 
$\ln k_{\perp}^2\sim \sqrt{\ln{s\over m^2}}$, 
but such terms do not spoil the ``power counting'' in $(\ln s)^n$). 
This assumption is not true for the diagrams leading to the renormalization
of the coupling constant since they are UV divergent.  However, the
structure of these diagrams is very simple - they are 1-loop vertex and 
self-energy corrections, and after the subtraction of the ${1\over 4-d}$ 
poles the integrals over $k_{\perp}^2$ are bound with $\mu^2$. In our case, 
the structure of 
the integrals over the transverse 
momenta is more complicated due to the mixture of the IR and UV divergencies.
There are individual diagrams where the upper
cutoff in the integrals over the $k_{\perp}$ is the energy $\sqrt{s}$ rather
than ${1\over a}$ or the normalization scale $\mu$.  Such contributions would 
give the additional non-BFKL $\ln s$ terms coming from the
logarithmical integrals over the transverse momenta 
$\int{dk_{\perp}\over
k_{\perp}^2}$.  
Fortunately, we will see below that such terms cancel in the
sum of all diagrams and the remaining transverse integrals are cut by the
dipoles' sizes (or by the normalization point $\mu$).

 \section{\label{crsc} Cross section}

\subsection{Calculation of Feynman diagrams} 

\begin{figure}
\includegraphics{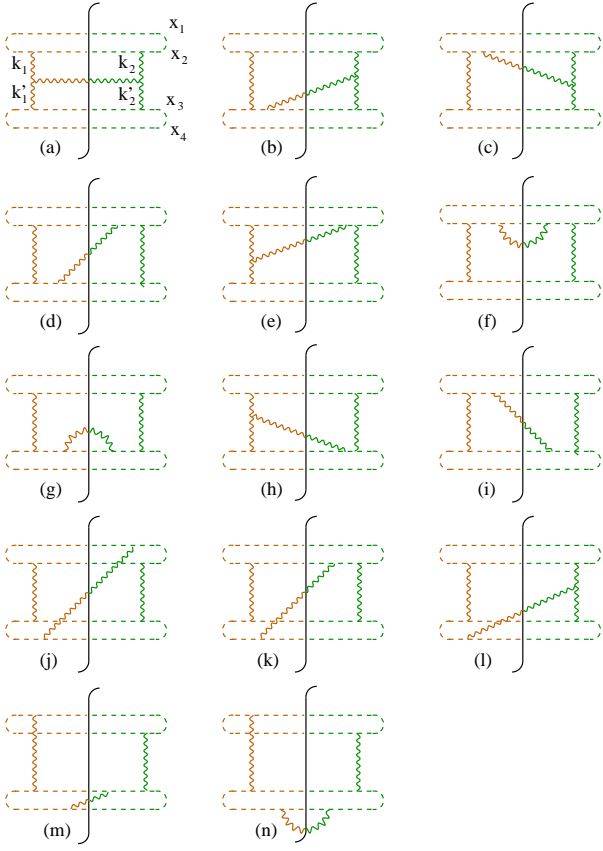}
\caption{\label{fig:dip7} Typical diagrams for the 
dipole-dipole cross section with 
the emission of a real gluon} 
\end{figure}

In the case of the cross section (\ref{opt1}), the structure of the diagrams 
is more transparent and we will present some details of this calculation. 
In this order, there are two types of Feynman diagrams: one with gluon
emission and one without. 
Typical diagrams of the first type are shown in Fig. \ref{fig:dip7}. 
The sum of these diagrams is proportional to the square of the finite-energy
Lipatov vertex $L^{\mu}(k_1,k'_1)L_{\mu}(k_2,k'_2)$. The Lipatov vertex  
\begin{eqnarray}
L(k_1,k'_1)&=&(k_1-k'_1)(ne)+\Big[2(k'_1n)+{(ne)\over(k_1e)}{k'}_1^2\Big]e
\nonumber\\
&-&
\Big[2(k_1e)+{(ne)\over(k'_1n)}k_1^2\Big]n
\label{lvertex}
\end{eqnarray}
is a sum of the emissions of the real
gluon with momentum $(k_1+k_1')$ from the left part of the diagrams 
in Fig. \ref{fig:dip6} in all possible ways. This vertex is gauge-invariant:
\begin{equation}
(k_1+k'_1)^{\mu}L(k_1,k'_1)=0
\end{equation}
At infinite energies $(ne)\rightarrow \infty$ the expression (\ref{lvertex})
reduces to the usual asymptotic form of Lipatov vertex (see e.g.
\cite{lobzor}).

\begin{figure}
 \includegraphics{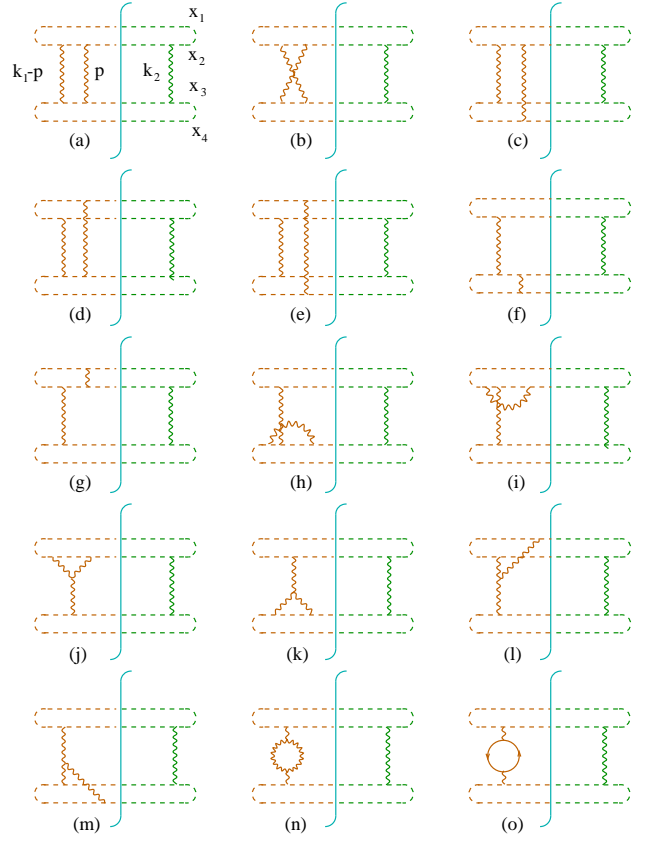}
\caption{\label{fig:dip8} Typical diagrams with 
an extra virtual gluon} 
\end{figure}

The sum of the diagrams in Fig. \ref{fig:dip7} has the form
\begin{eqnarray}
&&\hspace{0mm}
{4\over N_c^2-1}\varsigma(x_{1\perp},x_{2\perp};
x_{3\perp},x_{4\perp};\eta)_{\rm Fig.\ref{fig:dip7}}
\label{realg}\\
&&\hspace{0mm}
=~g^6N_c
\int\!{d^4 k_1d^4 k'_1 d^4 k_2d^4 k'_2
\over 128\pi^7k_1^2{k'}_1^2k_2^2{k'}_2^2}
\delta^{(4)}(k_1+k'_1-k_2-k'_2)
\nonumber\\
&&\times~
\delta(k_1+k'_1)^2\theta(k_1+k'_1)_0~\delta(k_1\cdot n)
\delta(k_2\cdot n)
\nonumber\\
&&\times~\delta(k'_1\cdot e)\delta(k'_2\cdot e)
(e^{-i(k_1,x_1)_{\perp}}-e^{-i(k_1,x_2)_{\perp}})~
\nonumber\\
&&\hspace{0mm}\times~
(e^{i(k_1,x_3)_{\perp}}-e^{i(k_1,x_4)_{\perp}})
~(e^{i(k_2,x_1)_{\perp}}-e^{i(k_2,x_2)_{\perp}})
\nonumber\\
&&\hspace{0mm}\times~
(e^{-i(k_2,x_3)_{\perp}}-e^{-i(k_2,x_4)_{\perp}})
\Big[-L(k_1,k'_1)\cdot L(k_2,k'_2)\Big]
~.
\nonumber
\end{eqnarray}
The explicit form of the product of two Lipatov vertices is
\begin{eqnarray}
&&\hspace{-4mm}-L(k_1,k'_1)\cdot L(k_2,k'_2)~
\label{liplip}\\
&&\hspace{-4mm}=~
{(ne)^3\over k_{1e}k'_{1n}}(k_1^2{k'}_2^2+k_2^2{k'}_1^2)
+2(k_1-k_2)^2(ne)^2
\nonumber\\
&&\hspace{-4mm}-~4({k'}_{1n}^2+k_{1e}^2)
-2(ne)\Big[
{k'_{1n}\over k_{1e}}
({k'}_1^2+{k'}_2^2)+{k_{1e}\over k'_{1n}}
(k_1^2+k_2^2)\Big]\nonumber\\
&&\hspace{-4mm}-~(ne)^2
\Big[{{k'}_1^2{k'}_2^2\over k_{1e}^2}
+{k_1^2k_2^2\over {k'}_{1n}^2}\Big]
\nonumber
\end{eqnarray}
where $k_{1e}\equiv (k_1\cdot e)$ and $k'_{1n}\equiv (k'_1\cdot n)$.
Apart from the diagrams with the emission of a real photon,  there
are diagrams with an extra virtual photon. If, for example, we take
the diagram in Fig. \ref{fig:dip2}h and insert an extra gluon line in the left
$(-)$ sector in all possible ways, we get the diagrams
shown in  Fig. \ref{fig:dip8}. We do not display the diagrams corresponding to
$\delta m$ or $\delta \epsilon$, see the discussion in Sect. \ref{lowest}.
The contribution of these diagrams vanishes here anyway since
the phase factors $e^{-i(2m+\epsilon){\cal T}}$ and $e^{i(2m+\epsilon){\cal
T}}$ coming from the left and from the right of the cut cancel out in the cross
section. Again, this leads to the replacement
of the color factors $-{1\over 2N_c}$ by $-{1\over
2N_c}-c_F=-{N_c\over 2}$ in the diagrams of the 
Fig.\ref{fig:dip7}m and Fig.\ref{fig:dip8}f-i type. 

The calculation yields:
\begin{eqnarray}
&&\hspace{0mm}
{4\over N_c^2-1}
\varsigma(x_{1\perp},x_{2\perp};
x_{3\perp},x_{4\perp};\eta)_{\rm Fig.\ref{fig:dip8}} 
\label{virg}\\
&&\hspace{0mm}
=~{g^4N_c\coth^2\eta\over 4\pi}
\int\!{d^2 k_{1\perp}d^2 k_{2\perp}\over 16\pi^4k_{1\perp}^2k_{2\perp}^2}
\nonumber\\
&&\hspace{0mm}\times~(e^{-i(k_1,x_1)_{\perp}}-e^{-i(k_1,x_2)_{\perp}})
(e^{i(k_1,x_3)_{\perp}}-e^{i(k_1,x_4)_{\perp}})\nonumber\\
&&\times
~(e^{i(k_2,x_1)_{\perp}}-e^{i(k_2,x_2)_{\perp}})
(e^{-i(k_2,x_3)_{\perp}}-e^{-i(k_2,x_4)_{\perp}})
\nonumber\\
&&\times~\Bigg[-\eta\coth\eta\!
\int\!{d^2 p_{\perp}\over 4\pi^2p_{\perp}^2} 
\Big({k_{1\perp}^2\over (k_1-p){\perp}^2}
+{k_{1\perp}^2\over (k_1-p){\perp}^2}\Big)
\nonumber\\
&&+~{1\over 4\pi}\Big(\ln{\mu^2\over k_{1\perp}^2}+
\ln{\mu^2\over k_{2\perp}^2}\Big)\Big({5\over 3}-{2n_f\over 3N_c}\Big)
\Bigg]
\nonumber
\end{eqnarray}
where the first ``gluon reggeization'' term comes from the 
diagrams in Fig. \ref{fig:dip8} a,b and the second from 
Fig. \ref{fig:dip8} n,o. As in the case of the amplitude, the diagrams
in Fig. \ref{fig:dip8} j-m vanish and the diagrams in Fig. \ref{fig:dip8} h,i
give rise to a pure divergency $\sim \int {dk_{\perp}\over k_{\perp}^2}$ 
that does not contribute in the framework of the dimensional regularization.
(Instead, it leads to the ``mixing'' of the IR and UV singularities, see the
discussion
 in Sect. \ref{sec:ricici}). Finally, the diagrams in Fig.
\ref{fig:dip8}c-e 
 cancel with the corresponding diagrams with an extra gluon
to the right of 
the cut. 

It is worth noting that the diagrams in Fig. \ref{fig:dip8} f,g 
should be regularized by finite ${\cal T}$ before calculation. 
After summation of all the relevant diagrams (e.g. Fig. \ref{fig:dip8} f,h 
and Fig. \ref{fig:dip7} g,m,n), 
the limit ${\cal T}\rightarrow\infty$ becomes regular. 

Performing the remaining integrations over longitudinal momenta in Eq.
\ref{realg} and the integration over $p$ in Eq. \ref{virg}, one
obtains
\begin{eqnarray}
&&\hspace{-3mm}\varsigma(x_{1\perp},x_{2\perp};x_{3\perp},x_{4\perp},\eta) 
\nonumber\\
&&\hspace{-3mm}=~{g^6N_c(N_c^2-1)\over 16\pi\tanh^2\eta} 
\int\!\dhd^2 k_1\dhd^2 k_2
\dhd^2 k'_1 \dhd^2 k'_2~
\label{mastereqs}\\
&&\hspace{-3mm}\times~
(e^{-i(k_1,x_1)}-e^{-i(k_1,x_2)})
(e^{-i(k'_1,x_3)}-e^{-i(k'_1,x_4)})
\nonumber\\
&&\hspace{-3mm}\times~
(e^{i(k_2,x_1)}-e^{i(k_2,x_2)})
(e^{i(k'_2,x_3)_{\perp}}-e^{i(k'_2,x_4)_{\perp}})
\nonumber\\
&&\hspace{-3mm}\times~
~\dbar^2(k_1+k'_1-k_2-k'_2)~
{a(k_i;\eta)\over(k_1+k'_1)^6}
\nonumber\\ 
&&\hspace{-3mm}+~
{g^2N_c(N_c^2-1)\over 16\pi\tanh^3\eta}\!\int\!
{\dhd^2 k_{1\perp}\dhd^2 k_2\over k_1^2k_2^2}
\nonumber\\
&&\hspace{-3mm}\times~
(e^{-i(k_1,x_1)}-e^{-i(k_1,x_2)})
(e^{i(k_1,x_3)}-e^{i(k_1,x_4)})\nonumber\\
&&\hspace{-3mm}\times~
(e^{i(k_2,x_1)}-e^{i(k_2,x_2)})
(e^{-i(k_2,x_3)}-e^{-i(k_2,x_4)})\nonumber\\
&&\hspace{-3mm}\times~
\Bigg[
-\eta \int\!\dhd^2 p~\Big\{{k_1^2\over p^2(k_1-p)^2}
+{k_2^2\over p^2(k_2-p)^2}\Big\}+
\nonumber\\
&&\hspace{-3mm}+~
{1\over 2\pi}\ln {\mu^2\over |k_1||k_2|}
\Big({5\over 3}-{n_f\over 2N_c}\Big)
+{1\over \pi}\ln \mu^2|x_{12}||x_{34}|\Bigg]
~.
\nonumber
\end{eqnarray}
Here 
\begin{equation}
a(k_i;\eta)=\sum_{j=1}^5a_j(k_i;\eta)
\end{equation}
where $a_1,a_2,a_3$, $a_4$ and $a_5$ correspond to the first, second, 
third, fourth and fifth terms in Eq. (\ref{liplip}), respectively. 
It should be mentioned that the fifth term in the r.h.s. of Eq. (\ref{liplip})
is IR divergent. After regularization, in addition to $a_5(k_i;\eta)$ it
produces 
the $\ln \mu^2x_{12}x_{34}$ term which forms ${11\over
3}N_c-{2\over 3}n_f$  
together with the contribution from the
self-energy
 diagrams in Fig. \ref{fig:dip8} n,o.

  The explicit form of the functions involved is:
\begin{equation}
a_i(k_1,k_2,k'_1,k'_2;\eta)~=~\Re A_i(k_1,k_2,k'_1,k'_2;\eta),
\end{equation}
where $A_i$ can be found in the Appendix B. We see that the ``unintegrated
cross section'' (\ref{mastereqs}) is given by the imaginary part of the
amplitude  (\ref{mastereqa}) (see the optical theorem for dipoles
(\ref{opt4})).

The cross section of dipole-dipole scattering is given by the integral
(\ref{opt0})
\begin{equation}
\sigma(\vec{a},\vec{b};\eta)~
=~\int\! d^2z~\varsigma(\vec{a}+\vec{z},\vec{z},\vec{b},\vec{0};\eta)
\label{crsect}
\end{equation}

\subsection{Asymptotics of the cross section}

As $\eta\rightarrow\infty$, the cross section (\ref{crsect}) reduces to
\begin{eqnarray}
&&\hspace{-2mm}
\sigma_{\rm asy}(\vec{a},\vec{b};s)=
\label{crscasy}\\
&&\hspace{-2mm}
=~g^4(N_c^2-1)
\int\!{\dhd^2 k\over k^2}{\dhd^2 k'\over{k'}^2} ~
4\sin^2{(k,a)\over 2}~\sin^2{(k',b)\over 2}\nonumber\\
&&\hspace{-2mm}\times~
\Bigg[
\Bigg(1+{g^2N_c\over 2\pi}~\Bigg\{
-\eta\! \int\!\dhd^2 p
{k^2\over
p^2(k-p)^2}
\nonumber\\
&&\hspace{-2mm}
+~{1\over 4\pi}\ln{\mu^2\over k^2}
\Big({5\over 3}-{2n_f\over 3N_c}\Big)
+{1\over 2\pi}\ln\mu^2 ab\Bigg\}\dbar(k+k')
\nonumber\\
&&\hspace{-2mm}+
~{g^2N_c\over 2\pi}\Bigg\{ {2\eta\over (k+k')^2}+
{\ln(k+k')^2/k^2\over (k+k')^2-k^2}\Big({2k^2\over {k'}^2}-1\Big)
\nonumber\\
&&\hspace{-2mm}+~
{\ln(k+k')^2/{k'}^2\over (k+k')^2-{k'}^2}
\Big({2{k'}^2\over k^2}-1\Big)-2{k^2+{k'}^2\over k^2{k'}^2}
\Bigg\}\Bigg]
+O(e^{-\eta})
\nonumber
\end{eqnarray} 
It is easy to see that the integrals over the transverse momenta converge
at $k_{\perp}^2\sim a^{-2}$. In the individual diagrams, the integrals
over $k_{\perp}^2$ diverge (which means that they would be cut by $s$ in
the corresponding exact expressions (\ref{crsect}) and (\ref{mastereqs})),
but in the sum of all diagrams such terms cancel (see the discussion in Sect.
\ref{sec:ricici}).   The structure of the asymptotic result Eq. (\ref{crscasy})
 is $\eta \sigma(a,b)$+$\sigma'(a,b)$.
The term $\sim\eta$ is the first iteration of the BFKL kernel. After
integration over $k$ and $k'$, it gives
the first term in the expansion of $\eta^{\alpha_s\chi(\nu)}$ in powers 
of $\alpha_s$. The second term may be called the ``dipole impact factor''.
Indeed, the standard representation of the BFKL asymptotics of the cross
section has the form
\begin{equation}
\sigma_{\rm asy}= \int\!{\dhd^2 k\over k^2}{\dhd^2 k'\over{k'}^2} 
I(a,k_{\perp})\eta^{K_{\rm BFKL}}I(b,k'_{\perp})
\end{equation}
where $K_{\rm BFKL}$ is the BFKL kernel (the LLA kernel + NLO BFKL
kernel $+...$). The kernel terms come from the central 
region of the rapidity, while the impact factors come from the fringes of 
the longitudinal integral
where the momentum of emitted gluon is almost collinear to $n$ or $e$.
 In this paper, we do not access the NLO BFKL terms \cite{nlobfkl} 
(those would
correspond to contributions to the amplitude $\sim\alpha_s^4\ln s$) so that 
all the non-LLA terms in Eq. (\ref{crscasy}) should be included 
in ``impact factors'' of the dipoles, namely, 
\begin{eqnarray}
&&\hspace{-3mm}I(b,k_{\perp})=\sin^2{(k,b)\over 2}+2\alpha_s k^2
\label{dipimp}\\
&&\hspace{-3mm}\times~
\int\!{\dhd^2 k'\over{k'}^2} 
\Bigg[ 
{\ln(k+k')^2/k^2\over (k+k')^2-k^2}\Big({2k^2\over {k'}^2}-1\Big)
-{2\over {k'}^2}\Bigg]\sin^2{(k',b)\over 2}
\nonumber
\end{eqnarray}
and similarly for $I(a,k'_{\perp})$. The impact factor(\ref{dipimp}) 
comes from the region of the longitudinal momenta in the emitted gluon 
close to $n$. Note that
the integral in the r.h.s of Eq. (\ref{dipimp}) is convergent and behaves like
$\ln k^2b^2$ for
 $k\rightarrow \infty$. Numerically, the $\alpha_s$ correction
to the impact
 factor (\ref{dipimp}) is quite significant - it is responsible 
for the difference between the dotted and dash-dot lines in Fig.
\ref{fig:numerico} below. (The NLO corrections to the virtual photon impact
factor calculated recently also appear to be big, see Refs.
\onlinecite{bartelsimp,fadinimp}).

\subsection{Unpolarized dipole-dipole scattering}

The ``unpolarized'' dipole-dipole scattering corresponds to the cross section
 (\ref{crsect}) averaged over the orientation of the dipoles : 
\begin{equation}
\sigma(a,b;s)~
=~\int\!{d\theta_a\over 2\pi}{d\theta_b\over 2\pi}\sigma(\vec{a},\vec{b};s)
\label{unpoldip}
\end{equation}
Note that in unpolarized $\gamma^\ast\gamma^\ast$
scattering the impact factor (\ref{impimpk}) depends only on $|\vec{a}|$ and
$|\vec{b}|$ so the cross section of the unpolarized 
$\gamma^\ast\gamma^\ast$ scattering (\ref{intro3}) is given by 
the ``unpolarized'' dipole-dipole cross section in Eq. (\ref{unpoldip}) 
integrated over the dipole sizes with weights
being impact factors
\begin{eqnarray}
&&\hspace{-4mm}
\sigma(a,b;\eta)~=~g^4(N_c^2-1)\coth^2\eta~
\label{crsca1}\\
&&\hspace{-4mm}
\times~\Bigg\{{g^2N_c\over 2\pi}
\int\!\dhd^2 k\dhd^2 k' ~
[1-J_0(ka)][1-J_0(k'b)]~
\nonumber\\
&&\hspace{-4mm}\times~{a_f(k,k';\eta)\over (k+k')^6}
~+~\int\!{\dhd^2 k\over k^4}[1-J_0(ka)][1-J_0(kb)]
\nonumber\\
&&\hspace{-4mm}
\times~\Bigg(1+{g^2N_c\over 2\pi}\Bigg[
-\eta\coth\eta\! \int\!\dhd^2 p~{k^2\over p^2(k-p)^2}
\nonumber\\
&&\hspace{-4mm}
+~{1\over 4\pi}
\Big({5\over 3}-{2n_f\over 3N_c}\Big)\ln{\mu^2\over k^2}
+{1\over 2\pi}\ln\mu^2ab
\Bigg]\Bigg)\Bigg\}
~.
\nonumber
\end{eqnarray} 
The explicit form of the function $a_f(k,k';\eta)$ 
is presented in the Appendix B.

\section{Numerical estimates of the dipole-dipole cross section}

\begin{figure}
\includegraphics{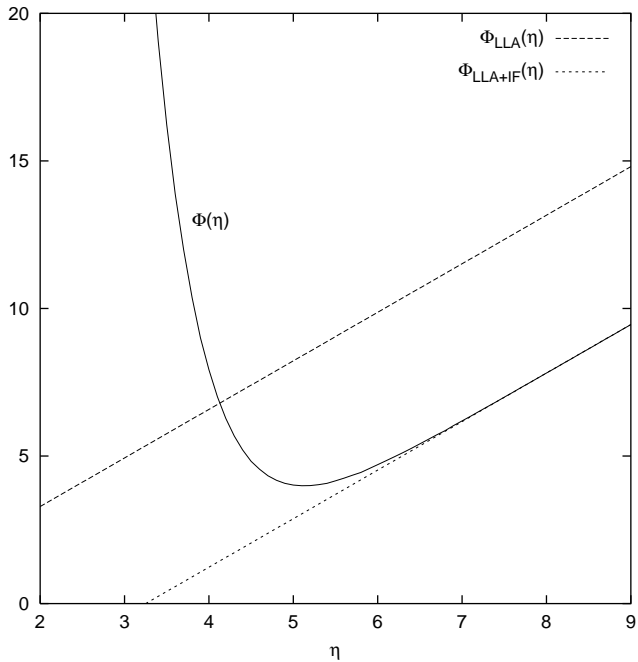}
\caption{\label{fig:numerico} Exact cross section and its BFKL asymptotics.
The solid line is Eq. (\ref{Phi}), the dotted one is $\Phi_{\rm LLA+IF}$ Eq.
(\ref{Phias}) and the dashed line is $\Phi_{\rm LLA}=1.65\eta$}
\end{figure}

 For numerical estimates, let us consider the unpolarized scattering of
equal-size dipoles. In this case we have only one-scale problem, so  
it is natural to take $\mu=a^{-1}$ where $a$ is the dipole size. 
The amplitude (\ref{crsca1}) reduces to 
\begin{equation}
\sigma(a,a;\eta)
~=~16\pi a^2\alpha^2_s(a)~\coth^2\eta~
\Big[1+6\alpha_s(a)\Phi(\eta)\Big]
\label{crsca2}
\end{equation}
where
\begin{eqnarray}
&&\hspace{-4mm}
\Phi(\eta)\label{Phi}\\
&&\hspace{-4mm}=~8\pi\!
\int\!\dhd^2 k\dhd^2 k' ~
[1-J_0(k)][1-J_0(k')]~
{a_f(k,k';\eta)\over (k+k')^6}
\nonumber\\
&&\hspace{-4mm}-
~8\pi\!\int\!{\dhd^2 k\over k^2}[1-J_0(k)]^2
~\Bigg[
\eta\coth\eta\! \int\!{\dhd^2 p\over p^2(k-p)^2}
+{\ln k\over 2\pi k^2}
\Bigg]
\nonumber
\end{eqnarray} 
where we have rescaled $k$ and $k'$ by $a$ and took $n_f=N_c=3$. 
This expression should be compared to the
asymptotical form (\ref{crscasy}) which gives 
\begin{equation} 
\hspace{-2mm}
\sigma_{\rm asy}(a,a;\eta)~=~16\pi\alpha_s^2(a)a^2~ 
\Big[1+6\alpha_s(a)\Phi_{\rm LLA+IF}(\eta)\Big] 
 \label{crscasy1}
\end{equation} 
where
\begin{eqnarray}
&&\hspace{-2mm} \Phi_{\rm LLA+IF}
=~8\pi\!
\int\!{\dhd^2 k\over k^2}{\dhd^2 k'\over{k'}^2} ~
[1-J_0(k)]
\label{Phias} \\
&&\hspace{-2mm}\times~[1-J_0(k')]~\Bigg\{ {2\eta\over (k+k')^2}+
{\ln(k+k')^2/k^2\over (k+k')^2-k^2}\Big({2k^2\over {k'}^2}-1\Big)
\nonumber\\
&&\hspace{-2mm}+~
{\ln(k+k')^2/{k'}^2\over (k+k')^2-{k'}^2}
\Big({2{k'}^2\over k^2}-1\Big)-2{k^2+{k'}^2\over k^2{k'}^2}
\Bigg\}
\nonumber\\
&&\hspace{-2mm}-~
8\pi\!
\int\!{\dhd^2 k\over k^4} ~
[1-J_0(k)]^2~\Bigg\{
\eta\! \int\!\dhd^2 p~{k^2\over p^2(k-p)^2}
+{\ln k\over 2\pi}
\Bigg\}
\nonumber
\end{eqnarray}
is a sum of the LLA term $\sim\eta$ and the impact factor term 
$\sim${\it const}. Numerically, 
$\Phi_{\rm LLA+IF}(\eta)\simeq 1.65(\eta-3.25)$ so  
\begin{equation} 
\sigma_{\rm asy}(a,a;\eta)~=~16\pi a^2\alpha^2_s(a) 
~\Big[1+9.88\alpha_s(a)(\eta -3.25)\Big] 
 \label{crscanum}
\end{equation} 
where $\eta$ corresponds to the LLA approximation and -3.251
 comes from the ``dipole  impact factor'' (\ref{dipimp}).
 We have not
calculated the next $\sim {1\over s}$ term in the asymptotic expansion
(\ref{crscanum}) 
but the fit to the Figure \ref{fig:numerico} suggests the coefficient of 
order of 100 in front of it.

On Figure \ref{fig:numerico}, we have depicted the comparison 
between the function $\Phi(\eta)$ describing the exact second-order cross
section, its asymptotics $\Phi_{\rm LLA+IF}(\eta)$ Eq. (\ref{Phias}) and 
the ``pure'' BFKL asymptotics $\Phi_{\rm LLA}(\eta)=1.64\eta$.
  
We have used the VEGAS algorithm of  the Monte-Carlo technique with 
10\% accuracy  to calculate the  function $\Phi(\eta)$.

We see that the BFKL asymptotics starts rather late, at $\eta\sim 5$, 
which translates into $\sqrt{s}\sim~10$ GeV for the scattering of dipoles 
with the $\rho$-meson size $a\sim 0.25 fm$. This is perhaps not surprising
since even within the (collinearly improved \cite{cia}) NLO BFKL approximation
itself
 the asymptotics starts relatively late, at $\eta\sim 6\div 8$ 
for the realistic dipoles \cite{trian}. It should be emphasized, however, that
these are two different theoretical questions relevant to the BFKL description
of experimental cross sections: (i) how good is the BFKL approximation to the
true high-energy asymptotics and (ii) when this true asymptotics starts making
sense. For the second-order dipole-dipole scattering the first question is
when the impact factor correction $\sim const$ can be neglected 
(in comparison to the LLA $\ln s$ term) whereas the
second question is when the ${1\over s}$ corrections disappear. 
In this paper, we
mainly address the second question and our conclusion is that the true
asymptotics (represented by $\Phi_{\rm LLA+IF}$ in this order) starts rather
late.

\section{Conclusions and outlook}

Our main result is that the scattering of dipoles is as suitable process
to study the energy behavior of QCD amplitudes as, for instance, the
scattering
 of virtual photons or oniums. As we discussed above, the 
reason why this conclusion could have been different is 
the infinite length of Wilson lines forming the dipoles 
that leads to the mixture of the IR and UV divergencies. 
There are individual diagrams where the upper cutoff 
in the integrals over the transverse momenta is $s$ rather than
${1\over a}$ or the normalization scale $\mu$.  Such  contributions would 
lead to the additional non-BFKL $\ln s$ terms coming from the integrals over 
the transverse momenta.  Fortunately in the sum of all
diagrams, such terms cancel so that the transverse integrals are cut 
with the dipoles sizes or normalization point $\mu$. The resulting 
dipole-dipole amplitude is an analytic function of the angle between 
the dipoles so that in principle, one can
get the BFKL kernel from the Euclidean calculation. 

The second conclusion is that for the dipole-dipole scattering the high-energy
asymptotics starts rather late, at $\eta\sim 5$. Of course, it
would be of more interest to check this statement for the scattering of virtual
photons rather than dipoles, but the corresponding diagram for the 
photon-photon scattering contains 4 loops so one should not expect this
calculation to be performed in the nearest future. 
  
Our statement about the lateness of asymptotical behavior refers 
to the scattering of equal-size dipoles. In many cases 
such as the deep inelastic scattering (DIS) one of the dipoles is small. 
The conventional assumption is that the propagation amplitude for the
small-size dipole through a hadron
is  proportional to the ``area of the dipole'' $a^2$
multiplied by gluon parton density normalized at $\mu^2=1/a^2$. This formula 
is valid for the asymptotically small dipoles, but it is unknown for 
how small dipoles and at what energies it starts making sense. 
This is an important question since in the 
analysis of the DIS from a nucleus based on the non-linear equation 
for the dipole evolution \cite{npb96,yura,mabraun}
we assume that at the energy of a few GeV and size of order of
saturation scale \cite{GLR,muchu,lrmodel,mu99}
(2$\div$3  GeV for LHC) the dipole matrix
element between the nucleon states satisfies this approximate formula
$a^2G(a^{-2})$.
Since we cannot
perform model-independent calculations of the dipole-hadron amplitudes at 
present, we can consider the ``target'' being the second dipole 
of a size of a typical hadron. The amplitude then is the $a\ll b$ limit of 
Eqn. (\ref{crsca1}). For this DIS from a color dipole we can answer 
the question of at what size of the spectator dipole the $a^2G$ approximation
start making sense. The study is in progress.

\begin{acknowledgments}

The authors are grateful to Yu.V. Kovchegov and L. McLerran for discussions.
 This work was supported by contract
DE-AC05-84ER40150 under which the Southeastern Universities Research
Association (SURA) operates the Thomas Jefferson National Accelerator
Facility and DOE grant DE-FG02-97ER41028. 
\end{acknowledgments}

\appendix
\section{Virtual photon impact factor}
In the momentum
representation, the $\gamma^\ast$ impact factor for
the transverse photons with the initial polarization $e_A$ and final $e'_A$ 
is given by
\begin{eqnarray} 
I^A(k_{\perp})&=&{1\over 2}\!\int\! d\alpha d\beta~
{1\over k_{\perp}^2\bar{\alpha}\alpha-p_A^2\bar{\beta}\beta}
\label{impimp}\\
&&\times~
\Big\{(1-2\bar{\alpha}\alpha)(1-2\bar{\beta}\beta)(e_A,e'_A)k_{\perp}^2
\nonumber\\
&&\hspace{-3mm}+~4\bar{\alpha}\alpha\bar{\beta}\beta
\Big[2(e_A,k)(e'_A,k)-(e_A,e'_A)k_{\perp}^2\Big]\Big\}
\nonumber
\end{eqnarray} 
where $\bar{x}\equiv 1-x$. If we average over the transverse polarizations,
the last term drops so that
\begin{equation} 
I^A(k_{\perp})~=
\int d\alpha d\beta~
{(1-2\bar{\alpha}\alpha)(1-2\bar{\beta}\beta)k_{\perp}^2
\over 2(k_{\perp}^2\bar{\alpha}\alpha-p_A^2\bar{\beta}\beta)}
\label{impimpa}
\end{equation} 
In the coordinate representation, 
the impact factor (\ref{impimpa}) takes the form
\begin{eqnarray} 
&&I^A(x_{\perp})\equiv\int{d^2 k\over 4\pi^2}e^{i(k,x)_{\perp}}I^A(k_{\perp})
\nonumber\\
&&I^A(x_{\perp})=\int d\alpha d\beta~
{(1-2\bar{\alpha}\alpha)(1-2\bar{\beta}\beta)
\over 4\pi\bar{\alpha}\alpha}
\sqrt{-p_A^2x_{\perp}^2{\bar{\alpha}\alpha\over\bar{\beta}\beta}}
\nonumber\\
&&\times~\Bigg[2K_1
\left(\sqrt{-p_A^2x_{\perp}^2{\bar{\alpha}\alpha\over\bar{\beta}\beta}}\right)
+\sqrt{-p_A^2x_{\perp}^2{\bar{\alpha}\alpha\over\bar{\beta}\beta}}
\nonumber\\
&&\hspace{16mm}\times~K_2
\left(\sqrt{-p_A^2x_{\perp}^2{\bar{\alpha}\alpha\over\bar{\beta}\beta}}\right)
\Bigg]
\label{impimpk}
\end{eqnarray} 
where $K$ is the McDonald function.
Note that the spin-averaged impact factor (\ref{impimpk}) depends only on
$x_{\perp}^2$.

\section{Explicit form of the second-order amplitudes}

As we mentioned above, the amplitude is a sum of five functions coming
from the five terms in the product of two Lipatov vertices
\begin{equation}
A(k_{i\perp};\eta)=\sum_{i=1}^5A(k_{i\perp};\eta)
\end{equation}

The explicit form of the first function is
\begin{eqnarray} 
&&\hspace{-3mm}A_1(k_{i\perp})={\cal A}_1(r_1,r'_2))+{\cal A}_1(r'_1,r_2)),
\label{B2}\\
&&\hspace{-3mm}{\cal A}_1(r_i,r_j))={2\eta\coth\eta\over r_ir_j}+
{\cosh^2\eta\over 4r_ir_j\cosh^2\eta
-(1-r_i-r_j)^2} 
\nonumber\\ 
&&\hspace{-3mm}\times~\Bigg[~\Big(l_i-{2\eta\tanh\eta\over r_i}\Big)
{4r_i\sinh^2\eta +3+r_i-r_j\over
(1-r_i)\sinh^2\eta}\nonumber\\
&&\hspace{-3mm}+~ \Big(l_j-{2\eta\tanh\eta\over r_j}\Big)
{4r_j\sinh^2\eta +3+r_j-r_i\over (1-r_j)\sinh^2\eta}
\nonumber\\
&&\hspace{-3mm}+~
{4\eta(1+r_i+r_j)\over r_ir_j\sinh 2\eta}\Bigg]\nonumber\\
&&\hspace{-3mm}-~
{i\pi\coth\eta\over r_ir_j}\Big[{r_i r_j\over (1-r_i)(1-r_j)}-
{1/2\over \sqrt{r_ir'_j}-1+r_i+r_j}\Big]
\nonumber
\end{eqnarray}
where $r_i\equiv {k_{i\perp}^2\over (k_1+k'_1)_{\perp}^2}$ and
\begin{equation}
l_i={\ln\Big[-1-2\sinh^2\eta
\Big(r_i+\sqrt{r_i^2+{r_i\over\sinh^2\eta}+i\epsilon}
\Big)\Big]
\over\sqrt{r_i^2+{r_i\over\sinh^2\eta}+i\epsilon}}
~.
\end{equation}


The second function is
\begin{eqnarray}
&&\hspace{-5mm}A_2\label{adva}\\
&&\hspace{-5mm}=~-2r_t
{M(r_1,r'_1)-M(r_2,r'_1)-M(r_1,r'_2)+M(r_2,r'_2)
\over (r_1-r_2)(r'_1-r'_2)}
\nonumber
\end{eqnarray}
where
\begin{eqnarray}
&&\hspace{-5mm}M(r_i,r_j)\\ 
&&\hspace{-5mm}=~{1\over 4\cosh^2\eta r_ir_j-(1-r_i-r_j)^2}~
\Bigg[(2r_i\cosh^2\eta
 \nonumber\\ 
&&\hspace{-5mm}+~
1-r_i-r_j)l_i+
(2r_j\cosh^2\eta+ 1-~r_i-r_j)l_i
\nonumber\\ 
&&\hspace{-5mm}-~2\eta\sinh 2\eta
+i\pi\Big(2\cosh\eta+{1-r_i-r_j\over \sqrt{r_ir_j}}\Big)\sinh\eta
\Bigg]
\nonumber
\end{eqnarray}
and
$r_t\equiv {-t\over (k_1+k'_1)_{\perp}^2}=
{(k_1-k_2)_{\perp}^2\over (k_1+k'_1)_{\perp}^2}$. Note that it does not
contribute to the forward scattering and therefore to the total cross section.


The third and the fourth functions are given by
\begin{eqnarray}
A_3&&=~{4\tanh^2\eta\over (r_1-r_2)(r'_1-r'_2)}
\label{atri}\\ 
&&\times~\Big[(r_1+r'_1)M(r_1,r'_1)
-(r_1+r'_2)M(r_1,r'_2)
\nonumber\\ 
&&-~(r_2+r'_1)M(r_2,r'_1)
-(r_2+r'_2)M(r_2,r'_2)\Big]
\nonumber
\end{eqnarray}
%
and
\begin{eqnarray}
A_4&&
\label{ache}\\ 
=&-&{2\tanh\eta\over r_1-r_2}
\Big[
N(r'_1,r_1)+N(r'_2,r_1)-(r_1\leftrightarrow r_2)\Big]
\nonumber\\ 
&-&{2\tanh\eta\over r'_1-r'_2}\Big[N(r_1,r'_1)+N(r_2,r'_1)
-(r'_1\leftrightarrow r'_2)\Big]
\nonumber
\end{eqnarray}
where
\begin{eqnarray}
N(r_i,r_j)&=&{1\over 4r_ir_j\cosh^2\eta -(1-r_i-r_j)^2}\label{Ns} \\ 
&\times&~
\Bigg[{4r_i\cosh^2\eta-r_j\over r_i-1}(r_i\coth\eta-2\eta+i\pi)
\nonumber\\ 
&+&~2\eta-i\pi+(1+r_i-r_j)l_j\coth\eta \nonumber\\  
&+&
3i\pi\sqrt{r_i\cosh^2\eta\over r_i\sinh^2
\eta+1}-2 i\pi\sqrt{r_i\over
r_j}\cosh\eta \Bigg]
~.
\nonumber
\end{eqnarray}
%

%
%

%
Finally,
\begin{eqnarray}
&&\hspace{-10mm}
A_5~ 
=~-{1\over r_1r_2}\Big({1\over r'_1}+{1\over r'_2}\Big)
-{1\over r'_1r'_2}\Big({1\over r_1}+{1\over r_2}\Big)
\nonumber\\ 
&&\hspace{-10mm}+~{l_1r_2-l_2r_1\over r_1r_2(r_1-r_2)\sinh^2\eta}
+{l'_1r'_2-l'_2r'_1\over r'_1r'_2(r'_1-r'_2)\sinh^2\eta }
~.
\label{apyat}
\end{eqnarray}
All the functions $A_i$ are analytical functions of the energy
($\equiv\eta$). In the Euclidean region $\eta$ should be replaced with
 $i\theta$ where $\theta$ is the angle between 
the dipoles ($\equiv$ Wilson rectangles). It is easy to see that 
the functions $A_i$ become real after such a replacement.

In the case of dipole-dipole cross section we need the real parts 
of the functions (\ref{B2}) - (\ref{apyat}) at $k_1^2=k_2^2$ and 
${k'}_1^2={k'}_2^2$. After some algebra we get
\begin{eqnarray}
&&a_f(r,r';\eta)\\
&&=~f_1(r,r';\eta)+f_3(r,r';\eta)+f_4(r,r';\eta)+f_5(r,r';\eta)
\nonumber
\end{eqnarray}
Here $f_1(r,r';\eta)=\Re {\cal A}(r,r';\eta)$ and
\begin{eqnarray}
f_3&=&2\tanh^2\eta {\partial^2\over\partial r\partial r'}
(r+r')m(r,r';\eta)\label{efs}\\
f_4&=&-2\tanh\eta \Big[{\partial\over\partial r}n(r',r;\eta)+
{\partial\over \partial r'}n(r,r';\eta)\Big]\nonumber\\
f_5&=&\Re{l-r\partial l/\partial r\over 2r^2\sinh^2\eta} 
+ 
\Re{l-r'\partial l'/\partial r'\over 2{r'}^2\sinh^2\eta} 
-{r+r'\over r^2{r'}^2}
\nonumber
\end{eqnarray} 
where $m(r,r';\eta)\equiv \Re M(r,r';\eta)$ and 
$n(r,r';\eta)\equiv \Re N(r,r';\eta)$.

\widetext
\mbox{}


\section*{References}

\vspace{-5mm}
\begin{thebibliography}{99}

\bibitem{mobzor}
I. Balitsky, {\it ``High-Energy QCD and Wilson Lines''}, 
In *Shifman, M. (ed.): At the frontier of particle 
physics, vol. 2*, p. 1237-1342  (World Scientific, Singapore,2001)
[hep-ph/0101042] 

\bibitem{nacht}
O. Nachtmann,
{\it Annals Phys.}  {\bf 209}, 436 (1991).

\bibitem{kop}
B.Z Kopeliovich, I.L. Lapidus, and Al.B. Zamolodchikov,
{\it JETP Letters} {\bf 33}, 612 (1981);

\bibitem{npb96}
I. Balitsky, 
{\it Nucl. Phys.}  {\bf B463}, 99 (1996).

\bibitem{prdpl}
I. Balitsky, 
{\it Phys. Rev.} {\bf D60}, 014020 (1999).
{\it Phys. Lett.} {\bf B 518}, 235 (2001)

\bibitem{dosch}
H.G. Dosch, {\it ``Space-Time Picture of High-Energy Scattering''}, 
In *Shifman, M. (ed.): At the frontier of particle 
physics, vol. 2*, p. 1195-1236  (World Scientific, Singapore,2001)

\bibitem{mu94}
A.H. Mueller, 
{\it Nucl. Phys.}  {\bf B415}, 373 (1994);
 A.H. Mueller and Bimal Patel, 
{\it Nucl. Phys.}  {\bf B425}, 471 (1994).

\bibitem{nn}
N.N. Nikolaev and B.G. Zakharov,
{\it Phys. Lett.} {\bf B 332}, 184 (1994);
{\it Z. Phys.}  {\bf C64}, 631 (1994).

\bibitem{balbel}
I. Balitsky and A. Belitsky, 
{\it Nucl. Phys.}  {\bf B629}, 290 (2002).

\bibitem{zahed}
M. Rho, S.-J. Sin and I. Zahed, 
{\it Phys. Lett.} {\bf B 466}, 199 (1999).

\bibitem{pesch}
R.A. Janik and R. Peschanski, 
{\it Nucl. Phys.}  {\bf B565}, 193 (2000);
{\it Nucl. Phys.}  {\bf B586}, 163 (2000).

\bibitem{dodik}
E.V. Shuryak and I. Zahed, 
{\it Phys. Rev.} {\bf D62}, 085014 (2000).

\bibitem{meggio}
E. Meggiolaro,
{\it Nucl. Phys.}  {\bf B625}, 312 (2002).

\bibitem{pirner}
A.I. Shoshi, F.D. Steffen, H.G. Dosch, and H.J.Pirner,
{\it ``Confining QCDstrings, Casimir scaling, and a Euclidean
   approach to high-energy scattering''},
Preprint HD-THEP-02-21, Nov. 2002,
[hep-ph/0211287].

\bibitem{bfkl}
V.S. Fadin, E.A. Kuraev, and L.N. Lipatov,
{\it Phys. Lett.} {\bf B 60}, 50 (1975);
I.I. Balitsky and L.N. Lipatov,
{\it Sov. Journ. Nucl. Phys.} 
{\bf 28}, 822 (1978).

\bibitem{lobzor}
L.N. Lipatov, 
{\it Phys. Rept.} {\bf 286}, 131 (1997).



\bibitem{mcler02}
E. Ferreiro, E. Iancu, K. Itakura, and L. McLerran, 
{\it Nucl. Phys.}  {\bf A710}, 373 (2002).

\bibitem{wkov02}
A. Kovner and U. Wiedemann, 
{\it Phys. Rev.} {\bf D66}, 051502 (2002).


\bibitem{kozlevin}
M. Kozlov and E. Levin,
{\it ``QCD saturation and $\gamma^\ast\gamma^\ast$ scattering''},
Preprint DESY-02-206, Nov. 2002,
[hep-ph/0211348].


\bibitem{keld}
I. Balitsky and V.M. Braun,   
{\it Nucl. Phys.}  {\bf B361}, 93 (1991);
{\it Nucl. Phys.}  {\bf B380}, 51 (1992).

\bibitem{difope}
I. Balitsky, 
{\it ``Operator expansion for diffractive high-energy scattering''},
[hep-ph/9706411]. 


\bibitem{hquarks}
M. Fischler,
{\it Nucl. Phys.}  {\bf B129}, 157 (1977).

\bibitem{diplom}
I. Balitsky, 
{\it Sov. Journ. Nucl. Phys.} 
{\bf 27}, 579 (1978).

\bibitem{nlobfkl}
V.S. Fadin and L.N. Lipatov,
{\it Phys. Lett.} {\bf B 429}, 127 (1998);
G. Carnici and M. Ciafaloni,
{\it Phys. Lett.} {\bf B 430}, 349 (1998).

\bibitem{bartelsimp}
J. Bartels, D. Colferai, S. Gieseke, and A. Kyrieleis,
{\it ``NLO corrections to the photon impact factor: 
combining real and virtual corrections''},
Preprint DESY-02-114, Aug. 2002,[hep-ph/0208130].

\bibitem{fadinimp}
V.S. Fadin and D. Yu. Ivanov and M.I. Kotsky,
{\it ``On the calculation of the NLO virtual photon impact factor''},
Preprint BUDKERINP-2002-60, Oct. 2002,
[hep-ph/0210406].

\bibitem{cia}
M. Chiafaloni, D. Colferai, and J. P. Salam,
{\it Phys. Rev.} {\bf D60}, 114036 (1999).

\bibitem{trian}
D.N. Triantafyllopoulos,
{\it ``The energy dependence of the saturation 
  momentum from RG improved BFKL evolution''},\\
Preprint CU-TP-1071, Sept. 2002,
[hep-ph/0209121].

\bibitem{yura}
Yu.V. Kovchegov,  
{\it Phys. Rev.} {\bf D60}, 034008 (1999);
{\it Phys. Rev.} {\bf D61},074018 (2000).

\bibitem{mabraun}
M.A. Braun,
{\it Eur.Phys.J.}{\bf C16}, 337 (2000), 
[hep-ph/0001268];


\bibitem{GLR}
L.V. Gribov, E.M. Levin, and M.G. Ryskin, 
{\it Phys. Rept.} {\bf 100}, 1 (1983).

\bibitem{muchu}
A.H. Mueller and J.W. Qiu  
{\it Nucl. Phys.}  {\bf B268}, 427 (1986).

\bibitem{lrmodel}
L. McLerran and R. Venugopalan,  
{\it Phys. Rev.} {\bf D49}, 2233 (1994);
{\it Phys. Rev.} {\bf D49}, 3352 (1994).


\bibitem{mu99}
A.H. Mueller,
{\it Nucl. Phys.} {\bf B558}, 285 (1999).

\end{thebibliography}
\end{document}